# National Accounts as a Stock-Flow Consistent System
# Part 1: The Real Accounts

Matti Estola[1]

University of Eastern Finland, Faculty of Social Sciences,
P.O. Box 111, FIN-80101, Joensuu, Finland
E-mail: matti.estola@uef.fi

and

Kristian Vepsäläinen

OP Financial Group
E-mail: kristian.vepsalainen@op.fi

**Abstract**

The 2008 economic crisis was not forecastable by at that time existing models of macroeconomics. Thus macroeconomics needs new tools. We introduce a model based on National Accounts that shows how macroeconomic sectors are interconnected. These connections explain the spread of business cycles from one industry to another and from financial sector to the real economy. These lingages cannot be explained by General Equilibrium type of models. Our model describes the real part of National Accounts (NA) of an economy. The accounts are presented in the form of a money flow diagram between the following macro-sectors: Non-financial firms, financial firms, households, government, and rest of the world. The model contains all main items in NA and the corresponding simulation model creates time paths for 59 key macroeconomic quantities for an unlimited future. Finnish data of NA from time period 1975-2012 is used in calibrating the parameters of the model, and the model follows the historical data with sufficient accuracy. Our study serves as a basis for systems analytic macro-models that can explain the positive and negative feedbacks in the production system of an economy. These feedbacks are born from interactions between economic units and between real and financial markets.[2] **JEL** E01, E10.

---

[1]Corresponding author.

[2]Thanks to Yougui Wang, Pirkko Aulin-Ahmavaara, Arto Kokkinen, Olli Savela, Alia Dannenberg, Anna Dannenberg and an anonymous referee for useful comments.

# 1. Introduction

The economic crisis at 2008 could not be forecasted by existing macroeconomic tools. Kobayashi (2009) writes: "*In fact, the crisis we are currently experiencing may call for a change in the theoretical structure of macroeconomics. In my view, a macroeconomic approach that encompasses financial intermediaries and places them at the centre of its models is necessary. We need a new paradigm of economic thought*".

Leijonhuvud (2009) criticises the General Equilibrium (GE) framework as follows: "*The economy is a large complex dynamical system which is in large measure self-regulating. Its self-regulatory features are the negative feedback loops that we refer to as 'market mechanisms': excess demand for a good raises its price which in turn reduces its excess demand. ... If the economy is displaced not too far from equilibrium, market forces will bring it back; if displaced too far, they will be ineffective or may work perversely. ... Theories that assume that the economy is a stable general equilibrium system ... do not hold in general. The instabilities that such theories ignore are precisely those problems that should be the particular responsibility of macroeconomists.*"

Caballero (2010) stresses the interactions in macro-models: "*The recent financial crisis has damaged the reputation of macroeconomics, largely for its inability to predict the impending financial and economic crisis. ... up to now the insight-building mode of the periphery of macroeconomics has proven to be more useful than the macro-machine-building mode. ... the periphery gave us frameworks to understand phenomena such as speculative bubbles, leverage cycles, fire sales, liquidity runs ... we need to spend much more effort in understanding the topology of interactions in real economies.*"

Roncaglia (2011) suggests that macroeconomic balance sheets are important elements in future macroeconomic models: "*The economic crisis we are now experiencing may entail radical changes in ... macroeconomics ... the series of events leading to the present crisis has made it quite clear that there are no automatic rebalancing mechanisms: thus analysis of the macroeconomic balance sheets and their interactions constitutes an important element in reconstruction of macroeconomics. In itself, the analysis of macroeconomic balance sheets is compatible both with the mainstream theories and with the heterodox theories: the context within which it is applied determines the cause-and-effect links characterizing the various interpretations of the economic events... Much still remains to be done, but the foundations for new macroeconomics are already available*".

We agree with these authors. Our framework is based on the System of National Accounts (SNA) that is not questioned in any macroeconomic theory. The closest framework to ours is the Stock-Flow Consistent (SFC) macro models developed by Godley (1983) and Lavoie (1992), and their book at 2007 is a good reference for SFC models. According to Accoce & Mouakil (2005), to build an SFC model requires 3 steps: writing the matrices, counting the variables and the accounting identities issued from the matrices, and defining for each unknown an equation: an accounting identity or a behavioral equation. SFC models are based on two tables: a balance sheet matrix of the stocks of real and financial assets of economic units in the beginning and ending moment of the accounting time, and a money flow matrix during the accounting time (Le Heron & Mouakil 2008).

Our model resembles those of Le Heron & Mouakil (2008) and E Silva & Dos Santos (2008) in many ways. However, we introduce an exact way how to create a money flow diagram of an account, and we work out the aggregation from micro units to macro sectors in detail. We also present diagrams for macro-level money and asset flow systems not given before. We show that all the circular flow diagrams in existing textbooks of macroeconomics are erroneous, and they do not correctly describe the principles of National Accounting. Our model, on the other hand, exactly corresponds to the principles of National Accounting, and thus we contribute in macroeconomic modelling that is in exact correspondence with NA. Our model simulates the historical data of Finnish NA reasonably well, and the model produces forecasts for 59 key macroeconomic quantities for an unlimited future. The model can be used in policy experiments and conditional forecasts of future scenarios.

Our study is organized as follows. First, we describe the principles of National Accounting. Then the principles of modelling flow systems in physics and engineering are introduced. Next, microeconomic units are defined as nodes in a money flow system, and the macro-sectors in an economy are aggregated from microeconomic units of equal kind. The SNA accounts are presented in the form of a money flow diagram between the macro-sectors, and similar diagrams are presented for property incomes, subsidies, and transfers. The correctness of our model is verified by the data of Finnish NA, and the simulations verify sufficient accuracy of the model with historical data.

## 2. Accounting Principles and the System of National Accounts[3]

The equation between the stock and the flow of a flowing quantity is:

$$F(t) = F(t_0) + \int_{t_0}^{t} F'(s)ds, \quad t > t_0, \qquad (1)$$

where $F(s)(unit)$ is the accumulating stock at time moment $s = t_0, t$, respectively, $F'(s) = dF/ds \,(unit/y)$ the corresponding net flow at time moment $s$, and by $s$ is denoted running time in units $y$, where $y$ is e.g. a year[4]. In physics, accumulated stocks of materials and time paths of particles are modelled. In money flows, the money stocks are deposit and loan capitals, and in real investment flows, the stocks are the fixed capitals of firms. Firms' bookkeeping, SFC-models, and SNA all are based on Eq. (1).

In firms' bookkeeping, opening Balance Sheet expresses the stocks of the assets and the liabilities of a firm, and the Profit and Loss Statement defines the revenues and costs of the firm during its accounting time unit. If function $F$ in Eq. (1) expresses the (stock of) net wealth of a firm, then $F(t_0)(€)$ is the net wealth in the beginning of the accounting time, $F'(s)(€/\Delta t)$ the profit during the accounting time $\Delta t = t - t_0$, and $F(t)(€)$ the net wealth of the firm at the ending moment of the accounting time.

The analogy between money and water flow system is demonstrated in Fig. 1, where the difference between inflow and outflow of water (money) during Δt corresponds to the Profit and Loss Statement of a firm, and the amount of water (money) in the vessel corresponds to the net wealth of the firm.

---

[3] In sections 2 and 3, there are many parallels with our earlier study Estola (2020).
[4] Measurement units are in parentheses after the quantities, see De Jong (1967).

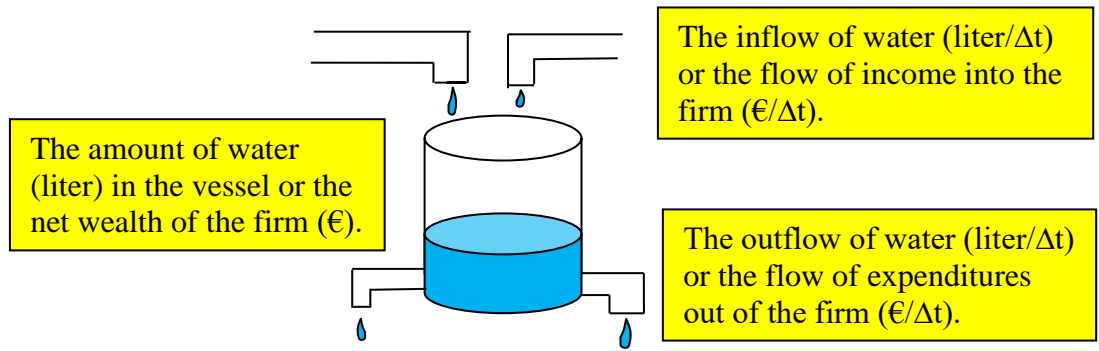

Figure 1. A water (money) flow system

The SNA and a money flow diagram are two ways to describe the real and the money flows between macro-economic sectors. In every trade, the expenditures of the buyer equal the gross revenues of the seller of which sales and product taxes are delivered to the state. Every bought good causes a money flow (its payment) in one direction, and a real flow (its delivery) in the opposite direction. The production process of an economy can thus be modelled on the basis of the money or the real flow system that have parallel events in both processes.

The nominal GDP of an economy is obtained by adding the value-added of all firms in the economy in a time unit at current prices. By the nominal GDP we can approximate the real GDP − the aggregate flow of production in the economy − as follows. Let all production volumes be measured in mass units, $kg$, and let the unit price of the product of firm $i$ be denoted by $p_i$ (€/$kg$) and the real value-added of the firm by $q_i(kg/y)$, where $y$ is a time unit. Let $n$ one-product firms exist in the economy. Then

$$GDP = \sum_{i=1}^{n} p_i q_i \approx \bar{p} \sum_{i=1}^{n} q_i \quad \Rightarrow \quad \frac{GDP}{\bar{p}} \approx \sum_{i=1}^{n} q_i,$$

where $\bar{p}$ (€/$kg$) is the average of $p_i, i = 1, ..., n$. An estimate for the real GDP, $GDP_R = \sum_i q_i(kg/y)$, is thus obtained by dividing the nominal $GDP$ by $\bar{p}$.

## 2.1. Modelling an Account

We have three analogous ways to present an account: 1) T -form, 2) Equation -form, and 3) Block diagram -form. These are in Fig. 2, where $R$ is revenues, $C$ costs, and $B$ the balancing item that makes both sides of the account equal; all these quantities have unit €/y. Block diagrams are common in physics and engineering in presenting flow systems, and we show that block diagrams can be applied in money flow systems too.

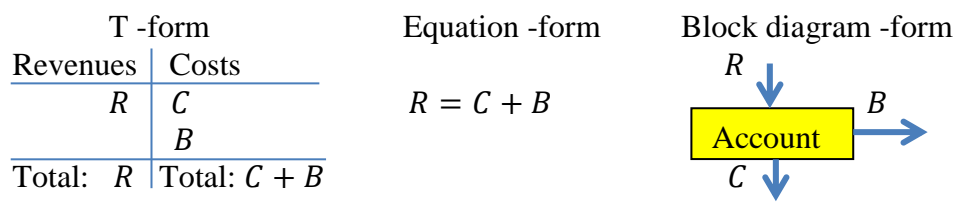

Figure 2. Three ways of presenting an account

Table 1. Accounting (transaction) matrix between macroeconomic sectors

|  | Households (HS) | Firms (FS) | Government (GS) | Sum |
|---|---|---|---|---|
| 1. Consumption | $-C$ | $C$ |  | 0 |
| 2. Government expenditures |  | $G$ | $-G$ | 0 |
| 3. Output |  | $[Y]$ |  |  |
| 4. Factor income | $W$ | $-W$ |  | 0 |
| 5. Taxes | $-T$ |  | $T$ | 0 |
| 6. Change in money stock | $-\triangle H$ |  | $\triangle H$ | 0 |
| Sum | 0 | 0 | 0 | 0 |

The transactions matrix in Table 1 is taken from Godley & Lavoie (2007 p. 60). This corresponds to the following money flow diagram between the sectors.

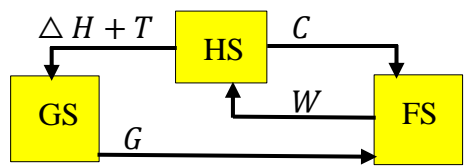

Figure 3. Money flow diagram corresponding to Table 1

The corresponding T-accounts and the system of equations are the following:

$$
\begin{array}{c|c}
\text{HS} & \\
\hline
W & C + T + \Delta H
\end{array}
\quad
\begin{array}{c|c}
\text{FS} & \\
\hline
C + G & W
\end{array}
\quad
\begin{array}{c|c}
\text{GS} & \\
\hline
T + \Delta H & G
\end{array}
\qquad
\begin{aligned}
HS: &\quad W = C + T + \triangle H, \\
FS: &\quad C + G = W, \\
GS: &\quad T + \triangle H = G.
\end{aligned}
$$

Thus, these four ways of presenting the system correspond to each other exactly.

### 3. The Principles of Modelling Flow Systems

Economic decision-making units can be described as nodes in a money flow system. In modelling a money flow system, we can apply similar principles as are applied in physics and engineering. Physical flow systems are water and other liquid flow systems, air and other gas flow systems, and electrical and solid material flow systems. In a flow system, we can apply the following modelling principles:

1) The conservation of mass, energy or momentum, or
2) Kirchhoff's current and voltage laws (Marshall 1978, 13, 16).

The conservation of mass means that material does not vanish. Material may change its form of existence, however; in heating water a part of it converts to steam and evaporates in the air. In a money flow system, we can assume that money does not vanish. If economic unit gets $Y(€)$, the unit either saves or consumes the money but does not destroy it. Thus $Y = S + C$ where $S(€)$ is savings and $C(€)$ consumption.

**Kirchhoff's Laws in Money Flow Systems**

*Current law:* In a money flow system, the sum of the money flows entering into a node (an economic unit) equals the sum of the money flows leaving the node, notice that saving too is considered as a money flow.

*Voltage law:* If some money enters in a money flow circuit, the same amount must flow out of the system for the amount of money ("energy") in the system to stay constant.

In the following, we assume the time unit to be one year and thus we study annual money flows in an economy.

## 4. Money Flows between Economic Units Originating from Production

### 4.1. Firm *i* as a Node in the Money Flow System of Production

Here firm *i* can be a financial or a non-financial firm, but later these two kinds of firms are treated separately. The opening balance sheet of firm *i* is:

*Assets (€)   Liabilities and net worth (€)*

| $A_{i0}$ | $\Psi_{i0} + NW_{i0}$ |
|---|---|

where the net worth $NW_{i0} = A_{i0} - \Psi_{i0}$ defines the wealth of the firm as the difference between its assets $A_{i0}$ and its liabilities $\Psi_{i0}$. In the first flow account of firm *i* – the production account (PA) – intermediate consumption is subtracted from the value of output ($\in P1$) of the firm (the official abbreviations of the items are in parentheses). This gives the value added of the firm. The gross value added ($\in B1G$) of firm *i* is: $GVA_i = value\ of\ output_i - H_i$, where $H_i$ ($\in P2$) is the intermediate consumption of firm *i* on the products of other domestic firms. The reason for subtracting intermediate consumption from the value of output is that in aggregating the gross value added of the economy, firms' productions would otherwise be added in several times.

The Net Value Added ($\in B1N$) is obtained by subtracting the consumption of fixed capital $P_i$ ($\in K1$) of the firm from $GVA_i$: $NVA_i = GVA_i - P_i$. The consumption of fixed capital corresponds to the depreciation bookings of the firm. In SNA, the consumption of fixed capital is not made on the basis of firms' depreciation bookings, however, but by the estimated consumption of firms' capital goods. In SNA, the valuation is made in gross and in net terms. We apply here net valuation where the consumption of fixed capital is taken into account.

Two kinds of pricing principles are applied in SNA. If production is valued at basic prices, taxes and subsidies on products are not included, and if production is valued at producers' prices, the taxes and subsidies on products are included in prices. For example, value added tax is collected as a fixed percentage of sales of goods (SNA 2009, 101-3). We value production at producers' prices.

Current transfers are aggregated into four items: 1) Net current taxes on income, wealth, etc. ($T_I \in D5$), 2) Social contributions and benefits other than social transfers in kind ($B_S \in D61 + D62,$), 3) Other current transfers ($N \in D7$), and 4) Capital transfers ($B_C \in D9$). Transfer is a transaction where an institutional unit provides a good, a service, or an asset to another unit without receiving from the latter any direct counterpart (SNA 2009, 161).

Social benefits can be divided in two classes: pensions and other social benefits. Other social benefits are collected either voluntarily or obligatory from households or firms, and are paid as transfers to the beneficiaries. Social transfers in kind contain, for example, final consumption expenditures undertaken by government on behalf of households (ibid, 167-178). With these assumptions, firm $i$ faces the following money flows with unit €/$year$.

*Revenues*: Households' ($C_{1i} \in P31$) and government's ($C_{2i} \in P31$) individual final consumption expenditures on the products of firm $i$, Government's collective final consumption expenditures ($G_i \in P32$) on the products of firm $i$, Firms' gross fixed capital formation ($I_i \in P5$) on the products of firm $i$, Sales of intermediate goods to domestic firms ($Z_i$), Exports of goods and services ($X_i \in P6$) of firm $i$, Property income ($PI_{Ri} \in D4$), Subsidies on production ($B_{Pi} = B_{P1i} + B_{P2i}$, $B_{Pi} \in D3$, $B_{P1i} \in D31$, $B_{P2i} \in D39$), Social contributions and benefits other than social transfers in kind ($B_{SiR}$), Other current transfers ($N_{Ri}$), and Capital transfers ($B_{CiR}$) received by firm $i$. $Z_i$ is an auxiliary variable, sub-index $R$ refers to revenues, $D3$ is Subsidies, $D31$ Subsidies on products, and $D39$ Other subsidies on production.

*Costs:* Expenditures on intermediate goods of other domestic firms ($H_i$), Employers' social contributions ($T_{Si} \in D12$), Taxes on production and imports ($T_{Pi} = T_{P1i} + T_{P2i}$, $T_{Pi} \in D2$, $T_{P1i} \in D21$, $T_{P2i} \in D29$), Gross wages and salaries ($W_i \in D11$), Current taxes on income, wealth, etc. ($T_{Ii}$), Own gross investment in fixed capital and changes in inventories ($I_{Oi}$), Social contributions and benefits other than social transfers in kind ($B_{SiP}$), Other current transfers ($N_{Pi}$), Capital transfers ($B_{CiP}$), Imports of goods and services ($M_i \in P7$), Net acquisitions of non-financial non-produced assets ($N_{Ain} \in K2$), and Property costs ($PI_{iP}$). Sub-index $P$ refers to payments, $D2$ to Taxes on production and imports, $D21$ to Taxes on products, and $D29$ to Other taxes on production.

Every firm does not face all these money flows, however. For example, if firm $i$ does not produce investment goods, then $I_i = 0$. In SNA in imports and exports, no distinction is made between consumption and investment goods, and between final and intermediate goods. We assume that all imports are carried out by the firms in the home country that sell the goods forward to domestic customers. Taxes on production and import contain, for example, value added tax.

The $NVA_i$ of firm $i$ is: $NVA_i = C_i + I_i + G_i + X_i + Z_i - M_i - H_i - P_i$, $C_i = C_{1i} + C_{2i}$. The imported goods sold by firm $i$ are subtracted from the market value of output to get the value of domestic production. In the generation of income account (GIA), the primary income components paid to the corresponding units are subtracted from the net value added. *Primary incomes* accrue to economic units due to their involvement in processes of production or ownership of assets that may be needed for production. This may contain compensation of employees, taxes on production and imports less subsidies, operating surplus or mixed income, and property income (SNA 2009, 131).

The balancing item in GIA is the operating surplus plus mixed income of firm $i$, $O_i \in B13N$, which is transferred to the allocation of primary income account APIA. *Mixed* *income* is the balancing item of unincorporated enterprises in the household sector that corresponds to remuneration for work carried out by the owner or members of his family including profits gained as entrepreneur (SNA 2009, 132).

In the APIA, $O_i$ is adjusted by property incomes $PI_{iR}$ and costs $PI_{iP}$. The balancing item is the balance of primary income of firm $i$, $BPI_i \in B5NT$. In the secondary distribution of income account SDIA, $BPI_i$ is adjusted by current transfers (SNA 2009, 159). This gives the disposable income of firm $i$, $DI_i \in B6N$, see Fig. 4.

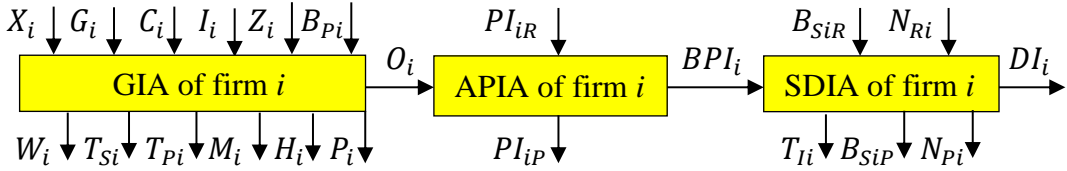

Figure 4. The GIA, the APIA, and the SDIA of firm $i$

Kirchhoff's current law is applied in every account (node) in Fig. 4, that is, the sum of the money flows entering equals the sum of the money flows leaving the account so that the balancing item is the difference. The equations for the accounts are:

$$O_i = X_i + G_i + C_i + Z_i + I_i + B_{Pi} - H_i - W_i - T_{Si} - T_{Pi} - M_i - P_i,$$

$$BPI_i = O_i + PI_{iR} - PI_{iP}, \quad DI_i = BPI_i + B_{SiR} + N_{Ri} - B_{SiP} - N_{Pi} - T_{Ii}.$$

Because firms do not participate in final consumption, the whole disposable income of firm $i$ is saved, $DI_i = S_i \in B8N$. Net saving links the use of disposable income account (UDIA) and the capital account (CA) of firm $i$, and here we omit the item *Adjustment for the change in pension entitlements (= D8)*, see SNA (2009, 181). The reason for omitting this marginal item is that pension contributions are included in social contributions (ibid, 168).

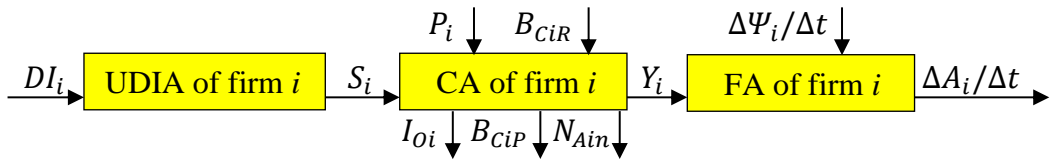

Figure 5. The UDIA, the CA, and the FA of firm $i$

The UDIA, the CA, and the financial account (FA) of firm $i$ are shown in Fig. 5. The flow of consumption of capital into CA can be explained in two ways: 1) CA measures the capital stock in net terms, and because investments increase the stock, the consumption of capital must be on the other side of the account. Thus, the capital stock of firm $i$ accumulates via the firm's net investment, $I_{Oi} - P_i$. Actually, firms do not pay anything of the consumption of their capital stock, and because $P_i$ enters as an "expenditure" item in the GIA of firm $i$, it must enter as a "revenue" item in some other account (CA) so that in the consolidation of the firm's accounts, $P_i$ cancels (see section 4.1.1). 2) The higher $P_i$ is, the smaller is $O_i$ and the less the firm pays current taxes on income that are calculated on the basis of $O_i$. Thus $P_i$ increases the savings of the firm and the firm benefits of its bookings of consumption of fixed capital even though this consumption speeds up the replacement of capital goods.

The net capital transfers received by firm $i$ are denoted by $B_{Cin} = B_{CiR} - B_{CiP}$ (through this study sub-index $n$ refers to net), $I_{Oi}$ is the (own) gross investments including changes in inventories, $Y_i (\in B9T)$ the net lending/borrowing, $\Delta \Psi_i / \Delta t$ the net issuing of financial liabilities, and $\Delta A_i / \Delta t$ the net acquiring of financial assets of the firm in the year; the last two are asset flows with $\Delta t = 1 \; year$. $N_{Ain}$ is the firm's net acquiring of non-financial non-produced assets If $N_{Ain} < 0$, firm $i$ has earned money by selling more than buying these assets. The equations of the accounts in Fig. 5 are:

$$S_i = DI_i, \quad Y_i = S_i + P_i + B_{Cin} - I_{Oi} - N_{Ain}, \quad \Delta A_i / \Delta t = Y_i + \Delta \Psi_i / \Delta t.$$

If $Y_i > 0$, firm $i$ has a surplus after its investments in fixed capital, and the firm can invest this money in financial assets or repay its loans. However, if $Y_i < 0$, this deficit must be financed by borrowing or issuing liabilities (notice that $\Delta A_i / \Delta t$ and $\Delta \Psi_i / \Delta t$ may be negative too). The acquiring of financial assets and liabilities is treated in a detailed way elsewhere (future publication Part 2: Financial Accounts).

Now, FA is the last flow account of firm $i$, and changes in stocks $\Delta A_i$, $\Delta \Psi_i$ with unit € are added, respectively, in the initial assets ($A_{i0}$) and liabilities ($\Psi_{i0}$) in the opening balance sheet of the firm. This gives the closing balance sheet of the firm at the end of the year. We omit here the *"Other changes in the volume of assets"* account and the *"Revaluation"* account that correct the amounts and the values of assets during the year due to non-intentional events, see SNA (2009, 237-255).

### 4.1.1. Consolidated Accounts of two Sectors of Firms

In consolidating aggregate sectors from individual economic units, the mutual money flows between the units cancel. For example, in the Non-Financial Firms' Sector (NFS) only those money flows remain visible NFS pays to or receives from other kinds of units. This is shown in Fig. 6 where Firm 1 and Firm 2 are consolidated into Firm 1+2, that receives all the incomes and pays all the expenditures of the two firms, and saves the rest. Firm 1 buys goods from Firm 2 by $R_3$.

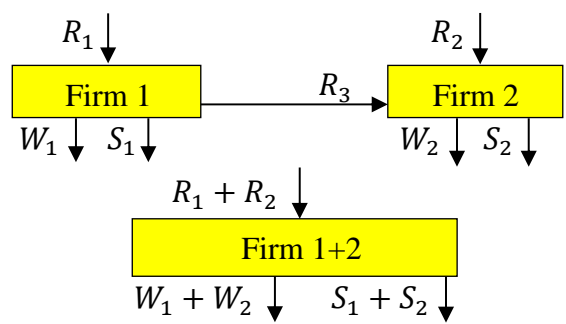

Figure 6. The consolidation of Firm 1 and Firm 2 into Firm 1+2

The consolidation of Firms 1 and 2 is made mathematically as follows. The income-expenditure equations of the firms are: $R_1 = R_3 + W_1 + S_1$ and $R_2 + R_3 = W_2 + S_2$. Adding the left and the right hand sides of the equations, the income-expenditure equation of Firm 1+2 becomes: $R = R_1 + R_2 = W_1 + W_2 + S_1 + S_2 = W + S$, where $R, W, S$ are the aggregate money flows faced by the consolidated firm. In this consolidation, the mutual money flow $R_3$ cancels.

By consolidating all non-financial firms in the home country into the sector NFS, the GIA, the APIA, and the SDIA of NFS become ($C_N = C_{1N} + C_{2N}$) as follows:

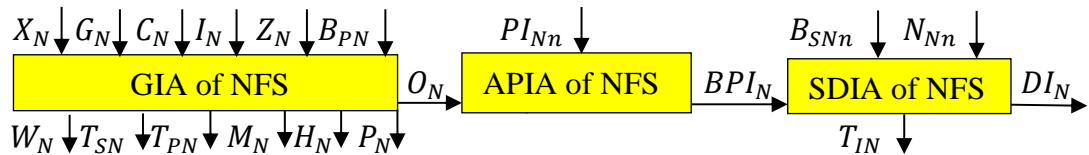

Figure 7. The GIA, the APIA, and the SDIA of NFS

The aggregate money flows faced in the accounts are explained in the Appendix. Similar aggregate quantities are defined for financial firms. The accounts for the Financial Firms' Sector (FFS) are as those for the NFS with the only deviation that the sub-index changes from N to F.

Next we show that our modeling is consistent with the Finnish National Accounts at year 2011 in million €/year. There $C_N + G_N + I_N + Z_N + X_N - M_N = 271298$ ($\in P1$). Further, $H_N = 166990$, $P_N = 21651$, $W_N = 50565$, $T_{SN} = 10875$, $T_{P2N} = 240$, and $B_{P2N} = 1222$, while $B_{P1N}, T_{P1N}$ are included in the value of production and are not reported in SNA (see section 5). These give $O_N = 22199$. Then, $PI_{NR} = 11430$ and $PI_{NP} = 22912$ give $BPI_N = 10717$. Still $B_{SNR} = B_{SNP} = 0$, $N_{NR} = 1666$, $N_{NP} = 1018$, and $T_{IN} = 4969$. Thus $DI_N = 6396$, as is correct.

The UDIA, the CA, and the FA of the NFS are:

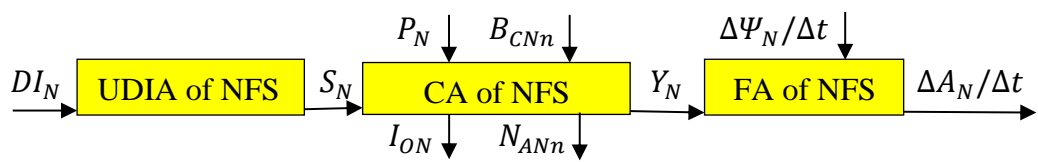

Figure 8. The UDIA, the CA, and the FA of NFS

In Finland at 2011, $DI_N = S_N = 6396$, $B_{CNR} = 255$, $B_{CNP} = 30$, $P_N = 21651$, $N_{ANn} = 111$, and $I_{ON} = 25016$. Thus $Y_N = 3367$ equals with that in Finnish NA.

The corresponding data for FFS in Finland at 2011 is: $C_F + G_F + I_F + Z_F + X_F - M_F = 9250$, $H_F = 4617$, $P_F = 450$, $W_F = 2295$, $T_{SF} = 473$, $T_{P2F} = 1$, and $B_{P2F} = 1$, while $B_{P1F}, T_{P1F}$ are not reported (see section 5). These give $O_F = 1415$. Then $PI_{FR} = 11618$ and $PI_{FP} = 11977$ give $BPI_F = 1056$. Further, $B_{SFR} = 1252$, $B_{SFP} = 1461$, $N_{FR} = 2245$, $N_{FP} = 2430$, and $T_{IF} = 638$ give $DI_F = 24$. Still $S_F = 92$ because $D8_{Fn} = 68$ (adjustment for the change in pension entitlements) is a marginal item that is excluded from our modeling. Then, $B_{CFR} = 49$, $B_{CFP} = 3$, $P_F = 450$, $N_{AFn} = 2$, and $I_{OF} = 356$. Thus $Y_F = 234$ equals with that in Finnish NA.

### 4.2. Household $j$ as a Node in the Money Flow System of Production

We omit here the opening balance sheet of household $j$ because it is similar to that of firm $i$, and we define only the relevant accounts for household $j$ for the accounting time (one year). Household $j$ may gain operating surplus plus mixed income due to working as an owner in an unincorporated firm (UF). In working as a market producer, household $j$ faces similar revenue and cost components as defined earlier for firm $i$ with the exception that the firms owned by households are assumed not to take part in international trade. These income and cost components are not repeated, however, but they appear in the GIA of household $j$. Household $j$ faces the following other revenues and expenditures with unit €/$year$.

*Revenues:* Gross wages and salaries ($W_{Tj}$), Employers' social contributions ($T_{SjR}$), Social contributions and benefits other than social transfers in kind ($B_{SjR}$), Other current transfers ($N_{jR}$), and Property incomes ($PI_{jR}$).

*Costs:* Expenditures on consumer goods ($C_{1j}$), Current taxes on income, wealth, etc. ($T_{Ij}$), Social contributions and benefits other than social transfers in kind ($B_{SjP}$), Other current transfers ($N_{Pj}$), Net acquisitions of non-financial non-produced assets ($N_{Ajn}$), and Property expenditures ($PI_{jP}$).

The GIA, the APIA, and the SDIA of household $j$ are then:

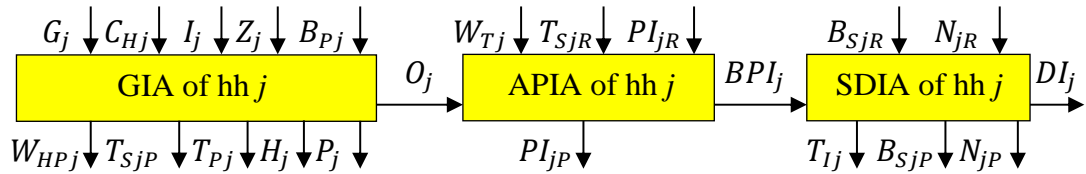

Figure 9. The GIA, the APIA, and the SDIA of household $j$

Now, $G_j$ is government's collective consumption expenditures on the products of the UFs owned by household $j$, $C_{Hj} = C_{1Hj} + C_{2Hj}$ is the individual final consumption of households and government on the products of the UFs owned by household $j$, and all the other factors in the GIA of household $j$ are as in the case of firm $i$. Even though households may operate as producers, most households earn only wage income. Thus $W_{HPj}$ is the gross wages paid by the UFs owned by household $j$ to their labourers, and $W_{Tj}$ is the possible wage income household $j$ receives as a labourer in a firm. $O_j$ is the operating surplus plus mixed income household $j$ receives as an entrepreneur, and $BPI_j$ is the balance of primary income after $O_j$ has been adjusted by wage and property incomes and costs. The disposable income of household $j$, $DI_j$, is obtained after social transfers have been taken into account.

The equations for the GIA, the APIA, and the SDIA for household $j$ are then:

$$O_j = G_j + C_{Hj} + I_j + Z_j + B_{Pj} - H_j - W_{HPj} - T_{SjP} - T_{Pj} - P_j,$$

$$BPI_j = O_j + W_{Tj} + T_{SjR} + PI_{jn}, \quad DI_j = BPI_j + B_{Sjn} + N_{jn} - T_{Ij}.$$

The capital goods of households correspond to the fixed capital of the UFs owned by households (SNA 2009, 82-83). Now, $S_j$ is the net saving, $P_j$ the consumption of fixed capital of the UFs owned by household $j$, $B_{CjR}$, $B_{CjP}$ are the capital transfers received and paid, $I_{Oj}$ the investments of the UFs owned by household $j$, $Y_j$ the net lending /borrowing, $\Delta\Psi_j/\Delta t$ the net acquiring of financial liabilities, and $\Delta A_j/\Delta t$ the net acquiring of financial assets of household $j$ in the year. The UDIA, the CA, and the FA of household $j$ are in Fig. 10, where $C_{1j}$ is the consumption expenditures.

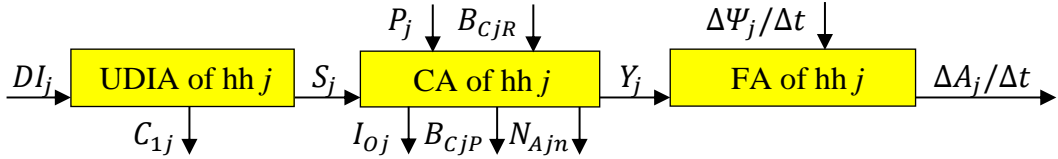

Figure 10. The UDIA, the CA and the FA of household $j$

Changes $\Delta\Psi_j$, $\Delta A_j$, respectively, are added in the opening balance sheet of household $j$ to get the closing one. By taking account the CA, the equation for FA in Fig. 10 is:

$$\Delta A_j/\Delta t = S_j + P_j + B_{Cjn} + \Delta\Psi_j/\Delta t - N_{Ajn} - I_{Oj}.$$

### 4.2.1. Consolidated Accounts of the Households' Sector (HS)

The Non-Profit Institutions Serving Households (NPISHs) – e.g. political parties and religious communities – are consolidated in the Households' sector (HS) because these units behave like households. The aggregate money flows faced by HS are described in the Appendix, and the consolidated accounts of sector HS+NPISH (shortly HS) are:

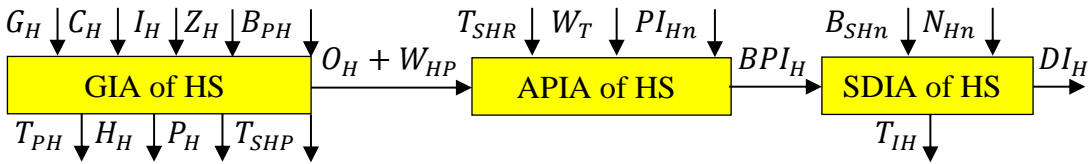

Figure 11. The GIA, the APIA, and the SDIA of HS

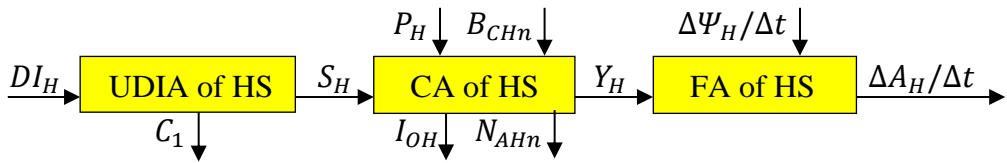

Figure 12. The UDIA, the CA, and the FA of HS

In the GIA, $C_H = C_{1H} + C_{2H}$ and HS pays gross wages $W_{HP}$ to its laborers. In the APIA, HS gets gross wages from other sectors as: $W_T = W_N + W_F + W_G + W_{RP} - W_{RR}$, see sections 4.3-4.4. The gross wage income of HS is thus: $W_{HR} = W_{HP} + W_T$.

The corresponding items in Finnish NA at 2011 in million €/year are: $C_H + G_H + I_H + Z_H = 35709 + 8297 = 44006$, $H_H = 13083 + 3808 = 16891$, $T_{P2H} = 3 + 0$, $T_{SHP} = 236 + 759 = 995$, $P_H = 7427 + 521 = 7948$, $B_{P2H} = 1565 + 0$, and $W_{HP} = 1048 + 3128 = 4176$, where the first numbers refer to HS and the latter to NPISH, and $B_{P1H}, T_{P1H}$ are not reported in Finnish NA, see section 5. Thus $O_H = 15477 + 81 = 15558$ equals with that in Finnish NA at 2011.

Now, $W_{HR} = 78677 + 0$, $T_{SHR} = 18224 + 0$, $PI_{HR} = 9632 + 530 = 10162$, and $PI_{HP} = 2212 + 55 = 2267$; thus $BPI_H = 119798 + 556 = 120354$. Also $B_{SHR} = 35336 + 0$, $B_{SHP} = 25265 + 0$, $N_{HR} = 1829 + 4478 = 6307$, $N_{HP} = 3487 + 481 = 3968$, and $T_{IH} = 25409 + 84 = 25493$. When the correcting item $D8_{Hn} = -68$ is taken into account, then $DI_H = 102802 - 68 + 4457 = 107191$, as is correct.

Further, $C_1 = 100464 + 5307 = 105771$ gives $S_H = 2270 - 850 = 1420$. Because $P_H = 7948$, $B_{CHR} = 235 + 230 = 465$, $B_{CHP} = 468 + 56 = 524$, $I_{OH} = 12896 + 528 = 13424$, and $N_{AHn} = 137 - 33 = 104$, then $Y_H = -3569 - 650 = -4219$ which is consistent with Finnish NA at 2011.

### 4.3. Government as a Node in the Money Flow System of Production

Government sector contains all legal entities established by political processes that have legislative, juridical or executive authority over other institutional units within a given area. Such entities are government, local governments, social security funds etc., see SNA (2009, 78-82). We assume, for simplicity, that government does not produce investment or export goods. Thus, government produces goods mostly for its own use or as transfer goods to other domestic units.

We omit here the opening balance sheet of government and start by defining the flow accounts for the government. From the market production of the government sector, consumers consume $C_G$ and the rest of it is used as intermediate consumption in firms ($Z_G$). The non-market output of the government sector is consumed partly by consumers (included in $C_G$) and mostly by the government sector itself, ($G_G$). Thus, government faces the following money flows with unit €/$year$.

*Revenues:* Sales of market goods to consumers ($C_G$) and firms ($Z_G$), Government's own consumption of public goods ($G_G$), Taxes on production and import ($T_{PGR}$), Current taxes on income, wealth etc. ($T_{IGR}$), Social contributions and benefits other than social transfers in kind ($B_{SGR}$), Other current transfers ($N_{GR}$), Capital transfers ($B_{CGR}$), and Property income ($PI_{GR}$).

*Costs:* Individual final consumption on goods ($C_2 = C_{N2} + C_{F2} + C_{H2}$), Collective final consumption on goods ($G = G_N + G_F + G_H + G_G \in P32$), Intermediate consumption ($H_G$), Subsidies to domestic and foreign producers ($B_{PGP}$), Employers' social contributions ($T_{SGP}$), Taxes on production and import ($T_{PGP}$), Current taxes on income, wealth etc. ($T_{IGP}$), Social contributions and benefits other than social transfers in kind ($B_{SGP}$), Other current transfers ($N_{GP}$), Capital transfers ($B_{CGP}$), Own investments in fixed capital ($I_{OG}$), Consumption of fixed capital ($P_G$), Net acquisitions of non-financial non-produced assets ($N_{AGn}$), and Property expenditures ($PI_{GP}$).

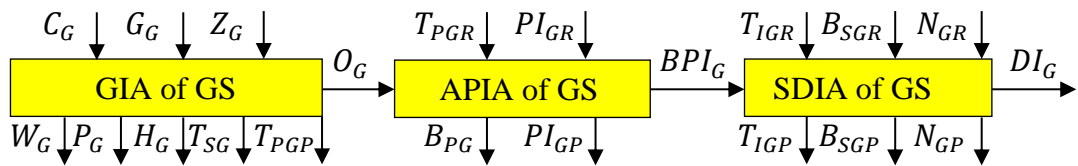

Figure 13. The GIA, the APIA, and the SDIA of GS

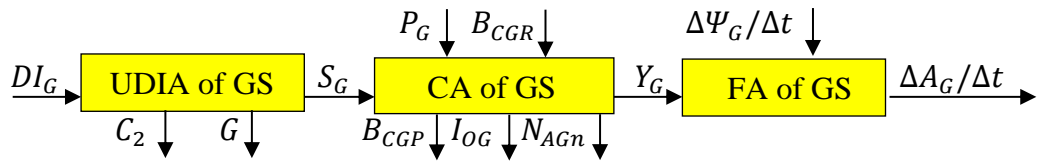

Figure 14. The UDIA, the CA, and the FA of GS

The accounts of GS are in Figs. 13 and 14. The net savings of government is denoted by $S_G$. If $S_G < 0$, government is a net borrower and vice versa. By taking account the APIA, the SDIA, and the CA, the equations for the UDIA and the FA of GS are:

$$S_G = O_G + T_{PGR} + T_{IGn} + PI_{Gn} + B_{SGn} + N_{Gn} - B_{PG} - C_2 - G,$$

$$\Delta A_G/\Delta t = S_G + P_G + B_{CGn} + \Delta \Psi_G/\Delta t - I_{OG} - N_{AGn}.$$

The data of Finnish NA at 2011 in million €/year are: $C_G + G_G + Z_G = 55816$, $H_G = 21418$, $T_{SGP} = 5905$, $T_{PG2P} = 4$, $P_G = 6532$, $W_G = 21544$, and $B_{PG2R} = 0$ while $B_{P1H}, T_{P1H}$ are not reported in Finnish NA, see section 5. These give $O_G = 413$ that equals with Finnish NA at 2011.

Now, $T_{PG1R} + T_{PG2R} = 26932 + 248 = 27180$, $B_{PG1P} + B_{PG2P} = 657 + 2067 = 2724$, $PI_{GR} = 7080$ and $PI_{GP} = 2896$ give $BPI_G = 29053$. Also, $B_{SGR} = 24037$, $B_{SGP} = 33876$, $N_{GR} = 27261$, $N_{GP} = 32173$, $T_{IGR} = 31209$, and $T_{IGP} = 109$. Then $DI_G = 45402$ that equals with Finnish NA at 2011.

Now $C_2 = 31229$ and $G = 15262$ give $S_G = -1089$, and $P_G = 6532$, $B_{CGR} = 835$, $B_{CGP} = 851$, $I_{OG} = 7486$, and $N_{AGn} = 3$ give $Y_G = -2056$, which is correct.

### 4.4. Rest of the World as a Node in the Money Flow System of Production

The capital stock of the rest of the world sector (RS) is not calculated in the SNA of the home country. The financial account of RS, on the other hand, defines the net lending position between the home country and RS. The home country and RS face the following mutual money flows with unit €/$year$ treated from the point of view of RS.

*Revenues*: Imports of home country ($M = M_N + M_F$), Wages and salaries ($W_{RR}$), Taxes on production and imports ($T_{PRR}$), Employers' social contributions ($T_{SRR}$), Social contributions and benefits other than social transfers in kind ($B_{SRR}$), Other current transfers ($N_{RR}$), Capital transfers $B_{CRR}$, and Property income $PI_{RR}$ from the home country.

*Costs:* Exports of home country ($X = X_N + X_F$), Wages and salaries ($W_{RP}$), Subsidies ($B_{PRP}$), Employers' social contributions ($T_{SRP}$), Social contributions and benefits other than social transfers in kind ($B_{SRP}$), Other current transfers ($N_{RP}$), Net acquisitions of non-financial non-produced assets ($N_{ARn}$), Capital transfers $B_{CRP}$, and property expenditures ($PI_{PR}$) to the home country.

The balancing item of the GIA of RS – the external balance of goods and services (EB) – measures the difference in sales of goods and services between the home country and RS. The balancing item of the consolidated account ASUA = APIA+SDIA+UDIA of the RS is the balance of payments ($BP$) between RS and the home country, and it equals with the disposable income $DI_R$ of RS.

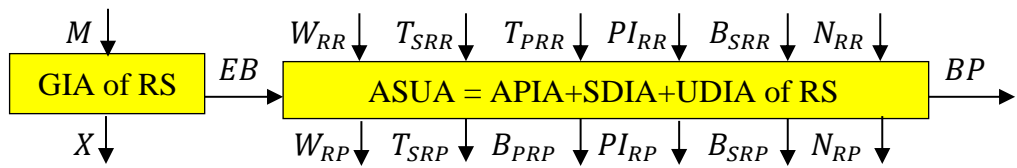

Figure 15. The GIA and the ASUA of RS

The corresponding equations for the GIA and the ASUA of RS are:

$$EB = M - X, \quad BP = M + W_{Rn} + T_{SRn} + PI_{Rn} + T_{PRR} + B_{SRn} + N_{Rn} - X - B_{PRP}.$$

Now, $BP = DI_R$ equals with the savings of RS, and after $BP$ is adjusted by net capital transfers $B_{CRn}$ and net acquisitions of non-financial non-produced assets $N_{ARn}$, we get the net lending/borrowing of RS to/from the home country, $Y_R$. The final flow account of RS is the FA that adds into $Y_R$ the net acquisition of liabilities of RS from the home country, $\Delta\Psi_R/\Delta t$. This gives the net acquisition of financial assets of RS from the home country, $\Delta A_R/\Delta t$. These accounts are shown in Fig. 16.

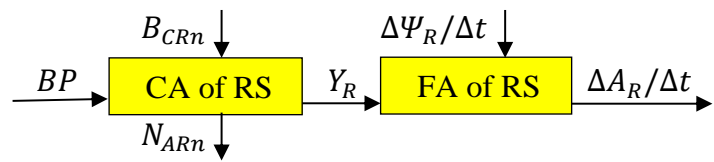

Figure 16. The CA and the FA of the rest of the world

By taking account the GIA, the APIA, and the SDIA, we get the equation for FA as:

$$\Delta A_R/\Delta t = BP + B_{CRn} + \Delta\Psi_R/\Delta t - N_{ARn}.$$

Adding changes $\Delta A_R$ and $\Delta \Psi_R$ in the opening balance sheet of RS we get the closing one. The Finnish NA at 2011 in million €/year are as follows. $M = 78768$, $X = 77093$ give $EB = 1675$. Further, $W_{RR} = 472$, $W_{RP} = 569$, $T_{SRR} = 116$, $T_{SRP} = 92$, $T_{PR1R} = 191$, $T_{PR2R} = 0$, $B_{PR1P} + B_{PR2P} = 51 + 721 = 772$, $PI_{RR} = 13613$, $PI_{RP} = 13851$, $B_{SRR} = 92 + 259 = 351$, $B_{SRP} = 116 + 246 = 362$, $N_{RR} = 3163$, and $N_{RP} = 1053$. These give $BP = 2882$. Now, $B_{CRR} = 10$, $B_{CRP} = 206$, and $N_{ARn} = 12$. These give $Y_R = 2674$, which is consistent with Finnish NA at 2011.

## 5. The Circular Money Flow Diagram Resulting from Production

The money flows between the consolidated macro-sectors originating from production are shown in Fig. 23. The accounts of the sectors are denoted in black/white by light grey boxes (in colours, yellow), the connections between the production system and the property income, financial asset, and the different subsidy markets are denoted by dark grey boxes (blue), and the accounts of RS are denoted by black boxes (red). Central bank has an essential role in property incomes and in financial asset markets, but these links are described in a detailed way elsewhere (Part 2).

In order to simplify Fig. 23, non-essential money flows within the sectors are eliminated by consolidating the APIA, SDIA, and UDIA of the sectors and renaming these as ASUA. The ASUAs of the sectors are in Figs. 17-19, while those of sectors FFS and RS are omitted because those of FFS and NFS are identical, and that of RS is in Fig. 15. The revenues and expenditures from intermediate goods are not described in Fig. 23 because at the aggregate level holds $H = Z$, and thus in the consolidation of the domestic sectors these money flows cancel.

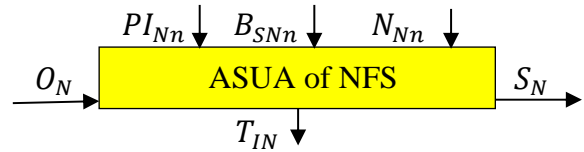

Figure 17. The consolidated ASUA of NFS

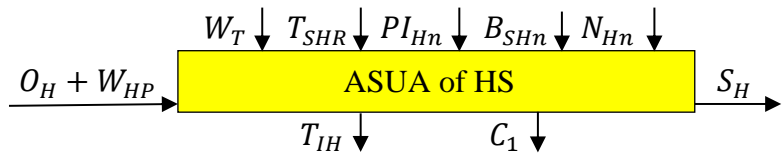

Figure 18. The consolidated ASUA of HS

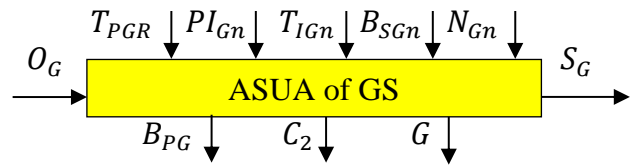

Figure 19. The consolidated ASUA of GS

The "markets", or different payment systems, are denoted by letter $M$ and are described in separate figures. In Fig. 20 is the system (market) of wages and salaries ($W \in D11$) paid and received in the production system in Fig. 23. According to Finnish NA at 2011, HS (including NPISH) and RS are the only sectors that receive wages and salaries together $78677 + 472 = 79149$ million €/year. The aggregate wages and salaries paid by the domestic sectors is 78580 million €/year: $W_G = 21544$, $W_F = 2295$, $W_N = 50565$, and $W_{HP} = 1048 + 3128 = 4176$, and RS pays the rest 569.

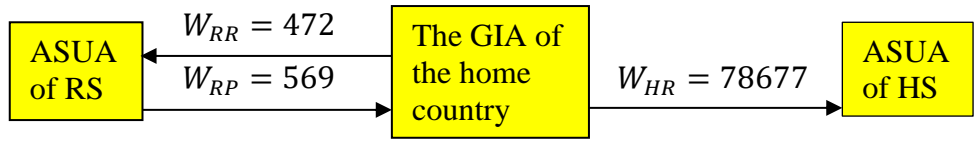

Figure 20. Wages and salaries in Figure 23

In Fig. 21 is the system of employers' social contributions ($T_S \in D12$) in Fig. 23. HS and RS receive these transfers, and in Finland at 2011, the aggregate employers' social contributions paid equals the aggregate received, 18340 million €/year.

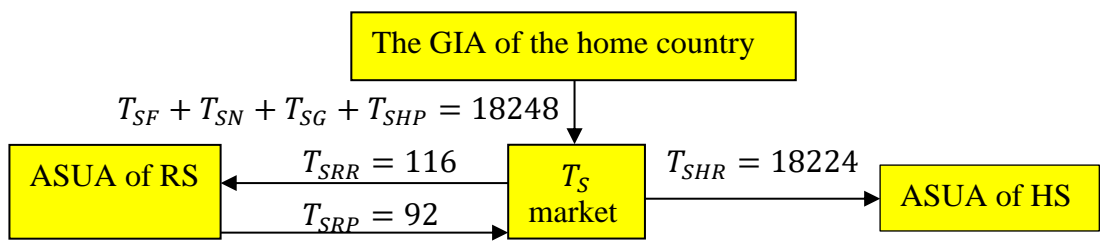

Figure 21. Employers' social contributions in Figure 23

Current taxes on income, wealth, etc. ($T_I \in D5$) is a zero sum system. The sectors pay and receive together 31209 million €/year to GS in Finland at 2011, see Fig. 23.

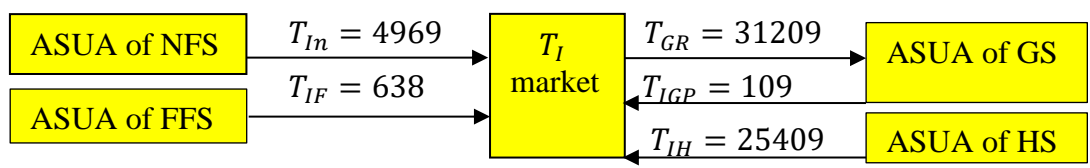

Figure 22. Current taxes on income and wealth, etc. in Figure 23

Figure 23. The money flow diagram originating from production

Property incomes and costs ($PI \in D4$) is a zero sum system. The latter number in HS in million €/year in Finland at 2011 refers to NPISHs, see Fig. 24.

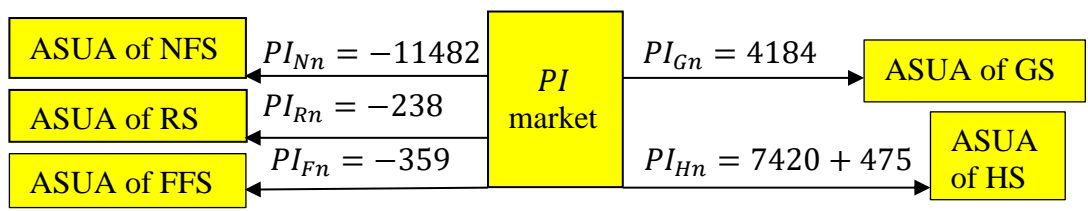

Figure 24. Net property incomes in Figure 23

Social contributions and benefits other than current transfers in kind ($B_S = D61 + D62$) is a zero sum system in Finland at 2011, where the sectors pay and receive, together, 60976 million €/year; see Fig. 25.

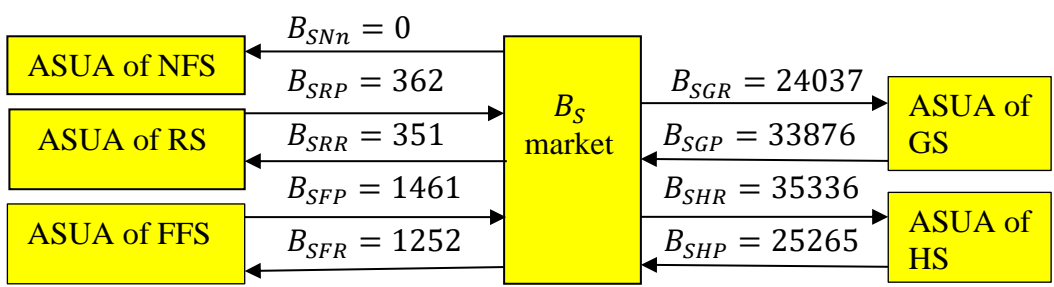

Figure 25. Social contributions and benefits other than social transfers in Figure 23

In Fig. 23, only RS and GS pay subsidies: $B_P = B_{P1} + B_{P2}$, $B_P \in D3$, $B_{P1} \in D31$, $B_{P2} \in D39$. D39 is a zero sum system in Finnish NA at 2011 where the sectors pay and receive, together, $(1565 + 0) + 1222 + 1 = 2788$ million €/year. The sums referring to HS + NPISHs are in parentheses and the rest refer to NFS and FFS. This equals with $B_{P2GP} + B_{P2RP} = 2788$. D31, on the other hand, is not a zero sum system. The 708 paid by RS and GS, together, increases the operating surpluses of the domestic sectors, and in Figure 23, $B_{PDR} = B_{P1HR} + B_{P1NR} + B_{P1FR} + B_{P2HR} + B_{P2NR} + B_{P2FR}$. In SNA it is not specified who receives subsidies D31 i.e. $B_{P1NR}, B_{P1FR}, B_{P1HR}$ are not reported. SNA reports only the sums of $B_{P1GP}$ and $B_{P1RP}$, see Fig. 26.

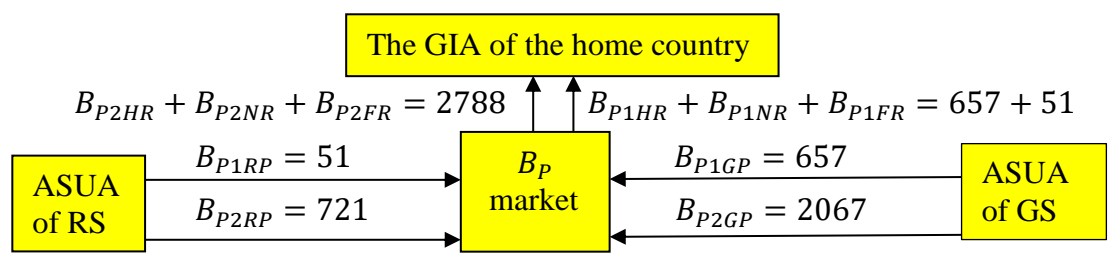

Figure 26. Subsidies in Figure 23

In Taxes on production and imports ($T_P \in D2$, $T_{P1} \in D21$, $T_{P2} \in D29$), domestic sectors pay and GS and RS receive the transfers. In Finland at 2011, D29 is a zero sum system where the sectors pay $T_{P2N} + T_{P2F} + T_{P2H} + T_{P2GP} = 248$ million €/year to GS. However, GS still receives 26932 million €/year as $D21$ in consumer prices. Taxes $D21$ are taken from sales and delivered to GS without specifying the sector that pays these taxes. RS receives, similarly, 191 million €/year (in the form of $D21$) from the sales of the home country. In Fig. 27, $T_{PR} = T_{P1R} + T_{P2R} = 191$, $T_{PGR} = 26932 + 248$, and $T_{PDP} = T_{P1N} + T_{P1F} + T_{P1H} + T_{P2N} + T_{P2F} + T_{P2H} + T_{P2GP} = 27371$.

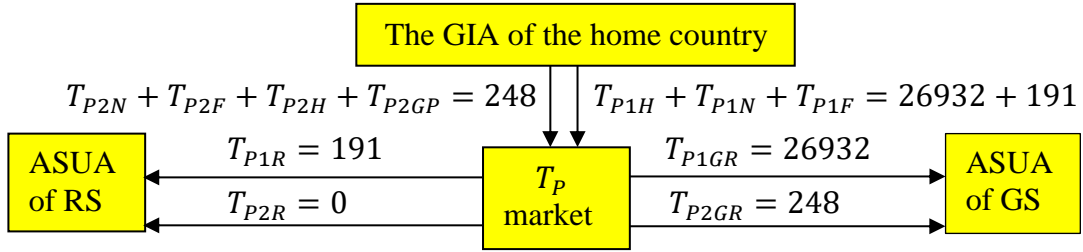

Figure 27. Taxes on production and imports in Figure 23

Other current transfers ($N \in D7$) is a zero sum system in Finland at 2011 so that the sectors pay and receive, together, 40642 million €/year, see Fig. 28.

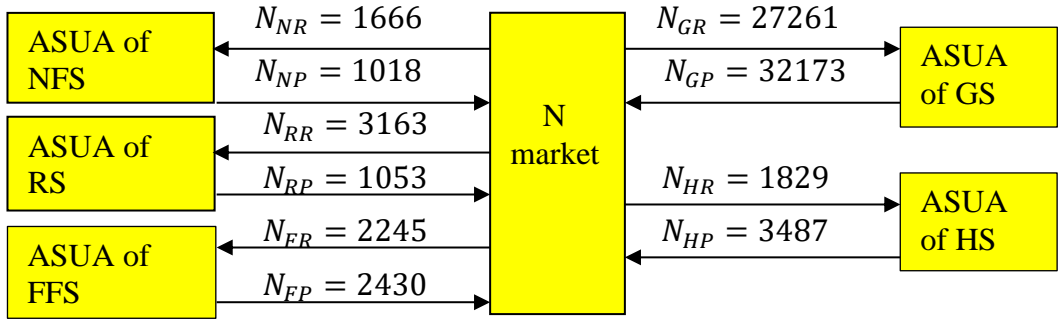

Figure 28. Other current transfers in Figure 23

Capital transfers ($B_C \in D9$) is a zero sum system in Finland at 2011 so that the sectors pay and receive, together, 1614 million €/year, see Fig. 29.

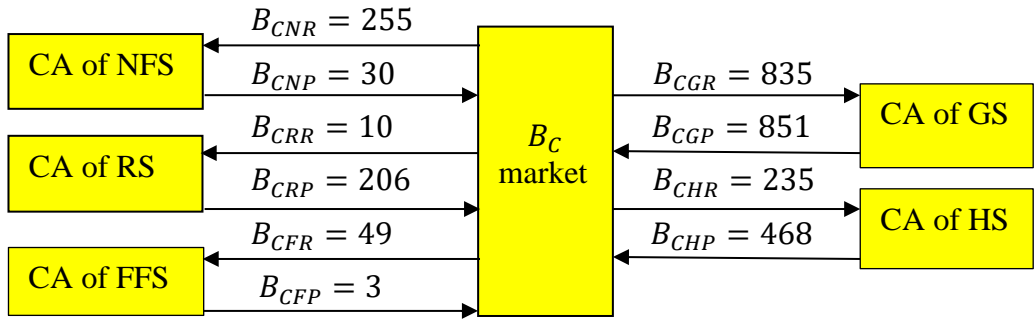

Figure 29. Capital transfers in Figure 23

Net acquisitions of non-financial non-produced assets ($N_A \in K2$) is a zero sum system in Finland at 2011. In HS, the latter number in million €/year refers to NPISH, Fig. 30.

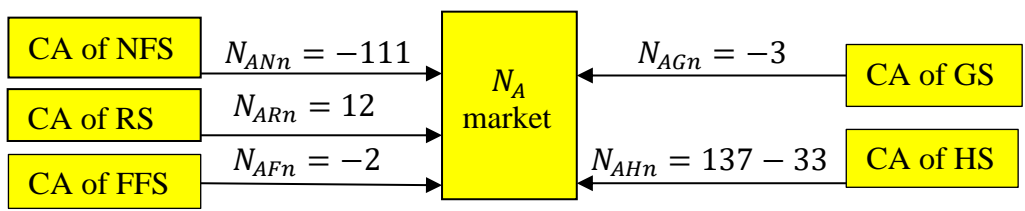

Figure 30. Net acquisitions of non-financial non-produced assets in Figure 23

Net lending/borrowing ($Y \in B9$) is a zero sum system in Finland at 2011, see Fig. 31.

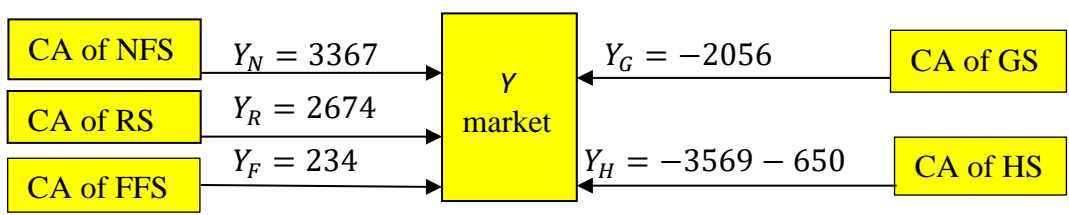

Figure 31. Net lending(+)/borrowing(−) in Figure 23

In Fig. 32 is a typical money-flow diagram presented in economics textbooks, see Cooper & John (2014). The reader can see that Figs. 23 and 32 essentially deviate from each other and that Fig. 32 contains only a fraction of all money flows registered in National Accounts and seen in Fig. 23. No economics textbook exists where in the circular flow diagram one economic sector - like the non-financial firms' sector - is divided into different accounts, like GIA, APIA, and CA, as is done in SNA. This disaggregation is essential for understanding the money flows that are internal within one sector and those between different sectors. For example, savings do not go from other sectors to the financial sector as in Fig. 32, but are internal money flows between accounts UDIA and CA in every sector. Another example is that the consumption of capital is not treated in Fig. 32 at all, which is an internal money flow from GIA to CA in every sector. Minor errors in Fig. 32 are that in it only financial institutions make investments, households do not pay taxes, government and firms do not save, etc. From Fig. 32 we cannot also derive the fundamental identities in National Accounting that are described in the following sections. An essential weakness of Fig. 32 is also that there the money flows are not derived by using any formal principle, while our block diagram presentations are identical with the corresponding T-accounts.

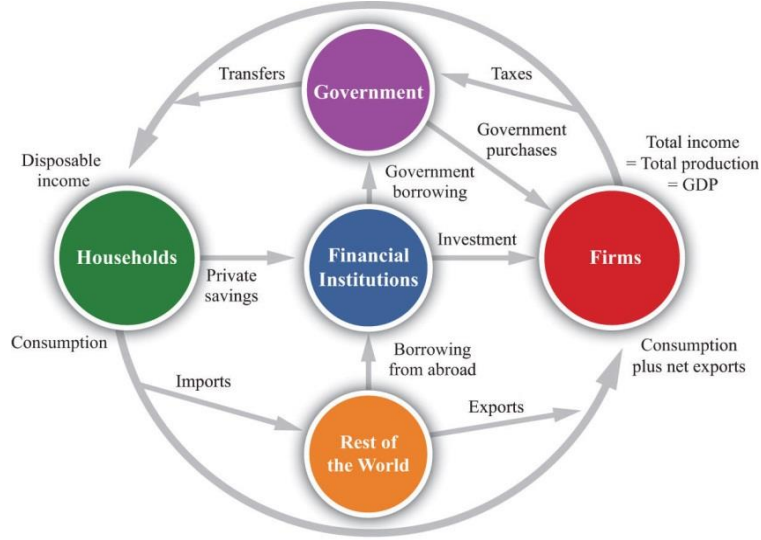

Figure 32. Macroeconomic money flow diagram (Cooper & John 2014)

### 5.1. Gross Domestic Product

In Fig. 23, box "The GIA of the home country" is defined by consolidating the GIAs of the domestic sectors NFS, FFS, HS, and GS. The corresponding equation is obtained by adding the inflows of money in the consolidated GIA of the domestic sectors and setting this equal to the sum of outflows of money from the consolidated GIA:

$$X_N + X_F + G_N + G_F + G_H + G_G + C_N + C_F + C_H + C_G + I_N + I_F + I_H + Z_N + Z_F + Z_H \\ + Z_G + B_{PN} + B_{PF} + B_{PH} \\ = O_N + O_F + O_G + O_H + W_N + W_F + W_{HP} + W_G + M_N + M_F + H_N \\ + H_F + H_G + H_H + P_N + P_F + P_H + P_G + T_{SN} + T_{SG} + T_{SF} + T_{SHP} \\ + T_{P1N} + T_{P1F} + T_{P1H} + T_{P2N} + T_{P2F} + T_{P2H} + T_{P2GP}. \qquad (2)$$

Because GS does not produce investment goods ($I_G = 0$) but only buys them ($I_{OG} > 0$), the aggregate investment in the economy is: $I = I_N + I_F + I_H = I_{ON} + I_{OF} + I_{OH} + I_{OG}$. The other aggregate quantities are: $X = X_N + X_F$, $M = M_N + M_F$, $G = G_N + G_F + G_H + G_G$, $C = C_1 + C_2 = C_N + C_F + C_H + C_G$, $Z = Z_N + Z_F + Z_G + Z_H$, $H = H_N + H_F + H_G + H_H$, $B_{PDR} = B_{P1NR} + B_{P1FR} + B_{P1HR} + B_{P2NR} + B_{P2FR} + B_{P2HR}$, $T_{SDP} = T_{SN} + T_{SF} + T_{SHP} + T_{SG}$, $T_{PDP} = T_{P1N} + T_{P1F} + T_{P1H} + T_{P2N} + T_{P2F} + T_{P2H} + T_{P2GP}$, $P = P_N + P_F + P_H + P_G$, $O = O_N + O_F + O_G + O_H$, and $W_{DP} = W_N + W_F + W_G + W_{HP}$.

By using the aggregate quantities, we can write Eq. (2) as

$$X + G + C + I + Z + B_{PDR} = O + W_{DP} + M + H + P + T_{SDP} + T_{PDP}$$

$$\Leftrightarrow C + I + G + (X - M) + Z - H = O + W_{DP} + T_{SDP} + T_{PDP} + P - B_{PDR}. \qquad (3)$$

On the left hand side of the latter form of Eq. (3) is the aggregate value of domestic production $C + I + G + (X - M) + Z$ including revenues from intermediate goods $Z$, and from this is subtracted the expenditures on intermediate goods $H$. Thus the left hand side of Eq. (3) equals with the Gross Value Added of the home country. Net exports are in parenthesis to separate international and domestic trade. In this consolidation, the revenues and expenditures on intermediate goods vanish because they are mutual money flows between the domestic sectors, see Fig. 6. Thus, $Z = H$, and then the left hand side of the latter form of Eq. (3) equals with GDP calculated on the basis of expenditures: $C + I + G + (X - M)$. On the right hand side of Eq. (3) is GDP calculated on the basis of income generated in production. Thus, the three ways of calculating GDP are obtained from Fig. 23, see SNA (2009, 332-3).

We check our modeling by the data of Finnish NA at 2011 in million €/year. There $X = 77093$, $M = 78768$, $G = 15262$, and $C = (100464 + 5307) + 31229 = 137000$, where the first two numbers refer to HS and NPISH, and GS takes part in Individual consumption (P31) by 31229. Further, $I = (12896 + 528) + 25016 + 356 + 7486 = 46282$, $T_{SDP} = (236 + 759) + 10875 + 473 + 5905 = 18248$, $P = (7427 + 521) + 21651 + 450 + 6532 = 36581$, $O = (15477 + 81) + 22199 + 1415 + 413 = 39585$, and $H = Z = (13083 + 3808) + 166990 + 4617 + 21418 = 209916$, where the numbers refer to (HS+NPISH), NFS, FFS, and GS. Still, $T_{PDP} = 27180 + 191 = 27371$, $B_{PDR} = 2724 + 772 = 3496$, and $W_{DP} = (1048 + 3128) + 50565 + 2295 + 21544 = 78580$.

We can now express the GDP of Finland at 2011 as: $C + G + I + (X - M) = 137000 + 15262 + 46282 + (77093 - 78768) = 196869$. To check that Eq. (3) is correct, we calculate the GDP on the basis of income: $O + W_{DP} + T_{SDP} + T_{PDP} + P - B_{PDR} = 39585 + 78580 + 18248 + 27371 + 36581 - 3496 = 196869$.

### 5.2. Disposable National Income

The second fundamental macro level identity in the system is obtained by adding the inflows of the UDIAs (disposable incomes) of all domestic sectors, and setting this equal with the sum of outflows of the corresponding UDIAs. This gives:

$$\begin{aligned} O_N + O_F + O_G + O_H + W_{HR} + PI_{Nn} + PI_{Fn} + PI_{Gn} + PI_{Hn} + B_{SNn} + B_{SFn} + B_{SHn} \\ + B_{SGn} + N_{Nn} + N_{Fn} + N_{Hn} + N_{Gn} + T_{SHR} + T_{PGR} + T_{IGR} \\ = T_{IN} + T_{IF} + T_{IH} + T_{IGP} + B_{PGP} + S_N + S_F + S_H + S_G + C + G \\ \Leftrightarrow \quad O + W_{HR} + T_{SHR} + T_{PGR} + N_{Dn} + B_{SDn} + PI_{Dn} - B_{PGP} = C + G + S_D, \quad (4) \end{aligned}$$

where $PI_{kn}$ is the net property income of sector $k$, $PI_{Dn} = PI_{Nn} + PI_{Fn} + PI_{Gn} + PI_{Hn} = -PI_{Rn}$, $B_{SDn} = B_{SNn} + B_{SFn} + B_{SHn} + B_{SGn}$, $S_D = S_N + S_F + S_G + S_H$, $T_{IGR} = T_{IN} + T_{IF} + T_{IH} + T_{IGP}$, and $N_{Dn} = N_{Nn} + N_{Fn} + N_{Hn} + N_{Gn}$. Taxes $T_I$ is on both sides of the equation and we can cancel it. Because $C + G$ is the aggregate final consumption and $S_D$ the aggregate saving of domestic sectors, the latter form of Eq. (4) shows that in the system money does not vanish; on the left hand side is total disposable income and on the right hand side total final consumption and saving in the country.

We check Eq. (4) by the data of Finnish NA at 2011 in million €/year. There $O = 39585$, $W_{HR} = 78677$, $T_{SHR} = 18224$, $T_{PGR} = 27180$, $N_{Dn} = -2110$, $B_{SDn} = 11$, $G = 15262$, $C = 137000$, $B_{PGP} = 2724$, $PI_{Dn} = 238$, and $S_D = (2270 - 850) +$

$6396 + 92 - 1089 = 6819$, where the numbers in parentheses refer to HS and NPISHs. These numbers fulfill Eq. (4).

## 6. Simulation results

We programmed the model so that starting from the total value of production, the GIA of the home country, the revenues are delivered to different sectors as is shown in Fig. 23. All money flows received by a "market" in Figs. 20-31 are first added together, and this money is then used in payments to other sectors as the figures describe. Historical correlations are applied in generating these money flows, and the program is made so that all the money received in the "market boxes" are used forward. Thus, every account in the system follows the "money does not disappear" principle. In Fig. 33 is shown the general structure of the model corresponding to Fig. 23, and the corresponding program file can be found in address: http://lchawk.arkku.net/tiedosto.mdl.

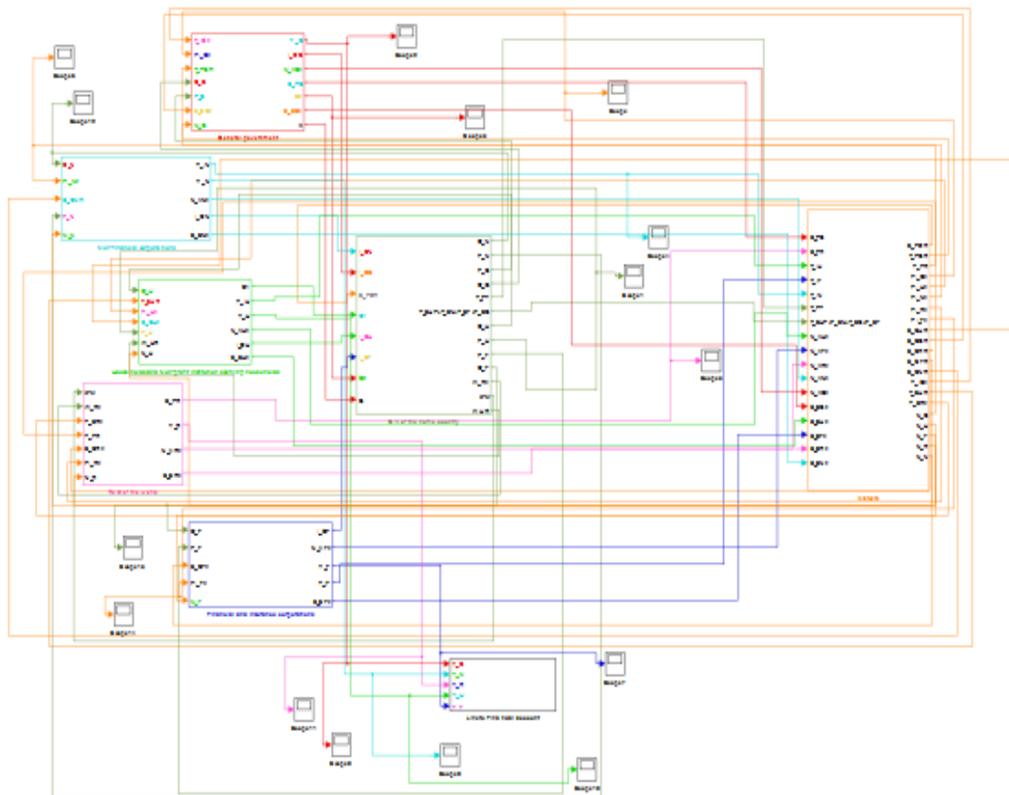

Figure 33. The structure of the simulation model

As an example of the form of the model, the block diagram corresponding to the ASUA account of sector HS is in Fig. 34.

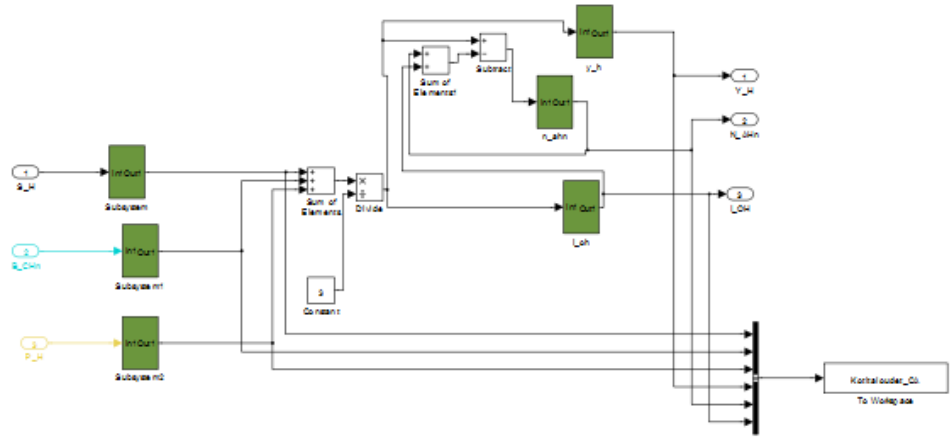

Figure 34. The ASUA account of sector HS

In the Appendix, in Fig. A1 is given a set of block diagrams in the model, and the simulation results concerning sector FFS are presented in Figs. A2-A7. Sector FFS was chosen for this demonstration because its variables fluctuate mostly of all the sectors and are thus the most difficult to simulate accurately.

## 7. Discussion

Macroeconomic theory needs a basis for modelling that allows quantitative analysis by using macro-level quantities, and the SNA is the only candidate giving data for this. Although Keynesian macro theory is based on the SNA, it applies only a small fraction of it. Our study shows how the principles of modelling flow processes in physics can be applied in money flow systems too such as the SNA. We modelled the links between expenditures and income generation in production and in property incomes, and also the links between the production system and financial market. The relations we presented for the macro-level income, transfer, and expenditure items gives a basis for new kind of macro models that describe the behaviour of an economy as a complex interacting system between economic units and real and financial markets. Hopefully this contributes in understanding and forecasting macro-level economic phenomena.

**Appendix**

NFS:

$X_N = \sum X_i$: Aggregate revenues of non-financial firms from exported goods,
$G_N = \sum G_i$: Aggregate government's collective consumption on the goods of non-financial firms,
$C_N = \sum C_i$: Aggregate households' and government's individual consumption on the goods of non-financial firms,
$I_N = \sum I_i$: Aggregate fixed capital formation on the goods of non-financial firms,
$Z_N = \sum Z_i$: Aggregate sales of intermediate goods by non-financial firms,
$H_N = \sum H_i$: Aggregate consumption of intermediate goods by non-financial firms,
$B_{PN} = \sum B_{Pi}$: Aggregate government subsidies to non-financial firms,
$O_N = \sum O_i$: Aggregate operating surplus plus mixed income of non-financial firms,
$W_N = \sum W_i$: Aggregate gross wages and salaries paid by non-financial firms,
$M_N = \sum M_i$: Aggregate expenditures of non-financial firms on imported goods,
$I_{ON} = \sum I_{Oi}$: Aggregate own fixed capital formation of non-financial firms,
$PI_{Nn} = \sum PI_{in}$: Aggregate net property incomes of non-financial firms,

$N_{Nn} = \sum N_{in}$: Aggregate net other current transfers received by non-financial firms,
$T_{IN} = \sum T_{Ii}$: Aggregate current taxes on income, wealth, etc. of non-financial firms,
$P_N = \sum P_i$: Aggregate depreciation of capital goods of non-financial firms,
$B_{CNn} = \sum B_{Cin}$: Aggregate net capital transfers received by non-financial firms,
$B_{SNn} = \sum B_{Sin}$: Aggregate net social contributions and social benefits other than social transfers in kind of non-financial firms,
$T_{PN} = \sum T_{Pi}$: Aggregate taxes on production and imports of non-financial firms,
$T_{SN} = \sum T_{Si}$: Aggregate employers' social contributions by non-financial firms.
$N_{ANn} = \sum N_{Ain}$: Net acquisitions of non-financial non-produced assets.

HS:

$G_H = \sum G_j$: Aggregate government's collective consumption on the goods of UFs owned by households,
$C_H = \sum C_{1Hj} + C_{2Hj}$: Aggregate households' and government's individual consumption on the goods of UFs owned by households,
$B_{PH} = \sum B_{Pj}$: Aggregate government subsidies to UFs owned by households,
$W_T = \sum W_{Tj}$: Households' aggregate gross wage income from working as an employee,
$W_{HP} = \sum W_{HPj}$: Aggregate gross wages paid by UFs owned by households,
$Z_H = \sum Z_j$: Aggregate sales of intermediate goods of UFs owned by households,
$H_H = \sum H_j$: Aggregate intermediate consumption of UFs owned by households,
$O_H = \sum O_j$: Aggregate gross operating surplus plus mixed income of UFs owned by households,
$I_H = \sum I_j$: Aggregate expenditures on investment goods produced by UFs owned by households,
$I_{OH} = \sum I_{Oj}$: Aggregate own investments in fixed capital of UFs owned by households,
$C_1 = \sum C_{1j}$: Aggregate households' final consumption expenditures,
$N_{Hn} = \sum N_{jn}$: Aggregate net other current transfers received by households,
$T_{PH} = \sum T_{Pj}$: Aggregate taxes on production and imports by UFs owned by households,
$T_{SHR} = \sum T_{SjR}$: Aggregate employers' social contributions received by households,
$T_{SHP} = \sum T_{SjP}$: Aggregate employers' social contributions paid by UFs owned by households,
$T_{IH} = \sum T_{Ij}$: Aggregate taxes on income, wealth, etc. paid by households,
$B_{SHn} = \sum B_{Sjn}$: Aggregate net social contributions and social benefits other than social transfers in kind received by households,
$S_H = \sum S_j$: Aggregate households' net savings,
$PI_{Hn} = \sum PI_{jn}$: Aggregate households' net property incomes,
$P_H = \sum P_j$: Aggregate consumption of fixed capital of UFs owned by households,
$B_{CHn} = \sum B_{Cjn}$: Aggregate net capital transfers received by UFs owned by households,
$N_{AHn} = \sum N_{Ajn}$: Net acquisitions of non-financial non-produced assets.

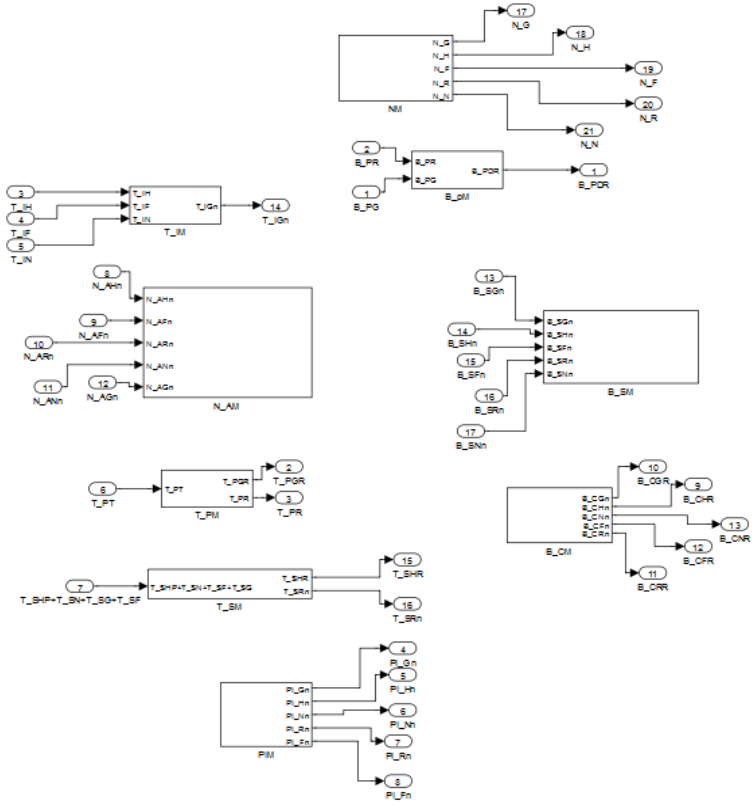

Figure A1. Some block diagrams in the model

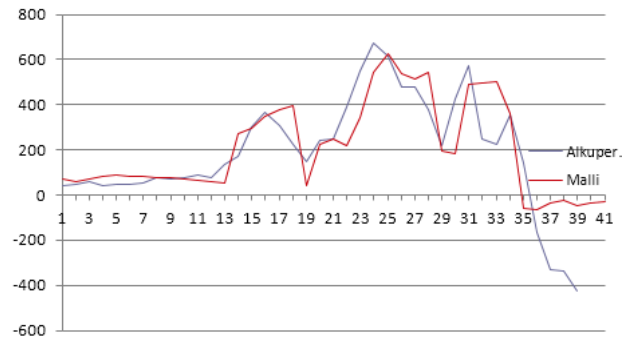

Figure A2. Simulation results for variable $B_{SF}$: the blue curve is original

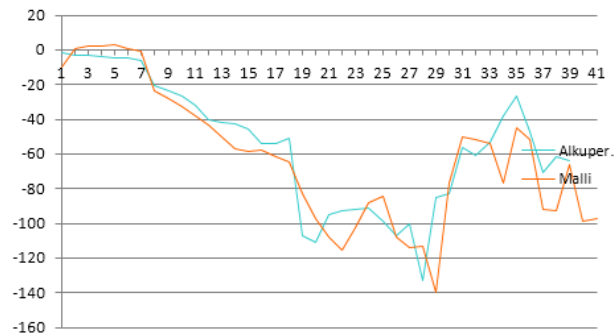

Figure A3. Simulation results for variable $N_F$: blue curve is original

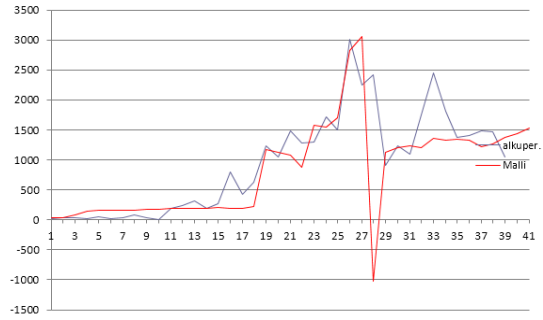

Figure A4. Simulation results for variable $O_F$: the blue curve is original

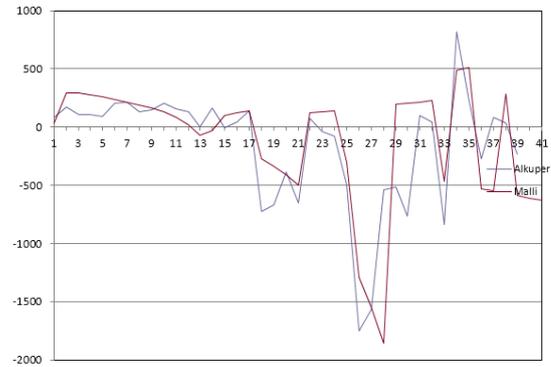

Figure A5. Simulation results for variable $PI_F$: the blue curve is original

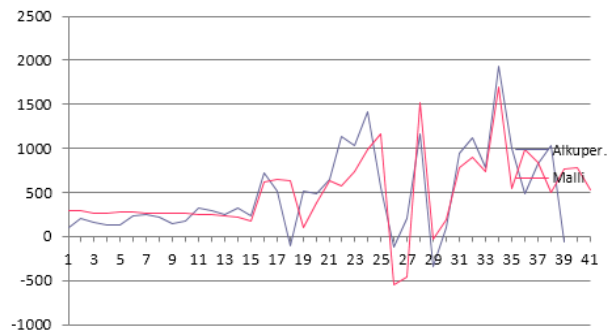

Figure A6. Simulation results for variable $S_F$: the blue curve is original

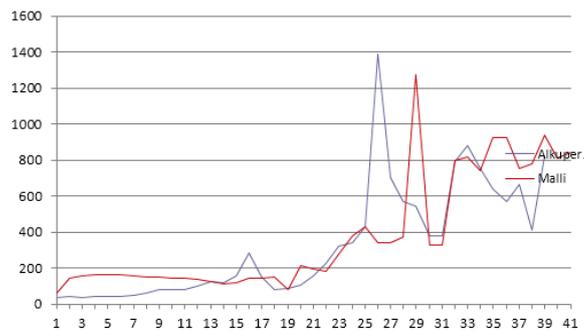

Figure A7. Simulation results for variable $T_{IF}$: the blue curve is original